\documentclass{article}
\usepackage{graphicx}
\usepackage[round]{natbib}
\usepackage{epsfig}
\title{
Predicting landfalling hurricane numbers from basin hurricane
numbers: statistical analysis and predictions
}

\author{
Stephen Jewson (RMS)\footnote{\emph{Correspondence email}: \texttt{stephen.jewson@rms.com}}\\
Thomas Laepple (AWI)\\
Adam O'Shay (RMS)\\
Jeremy Penzer (LSE)\\
Enrica Bellone (RMS)\\
Kechi Nzerem (RMS)\\
}

\begin{document}
\maketitle

\begin{abstract}
One possible method for predicting landfalling hurricane numbers
is to first predict the number of hurricanes in the basin and
then convert that prediction to a prediction of landfalling
hurricane numbers using an estimated proportion. Should this work
better than just predicting landfalling hurricane numbers
directly? We perform a basic statistical analysis of this question
in the context of a simple abstract model, and convert some previous predictions
of basin numbers into landfalling numbers.
\end{abstract}

\section{Introduction}

We are interested in trying to develop and compare methods for the
prediction of the distribution of the number of hurricanes that
might make landfall in the US in future years. One class of
possible methods that one might use involves first predicting the
number of hurricanes in the Atlantic basin, and then converting
that prediction to a prediction of landfalling numbers using some estimate
of the proportion that might make landfall. Is this class of \emph{indirect} methods
likely to work any better than simpler methods based on predicting
the number of landfalls directly? On the one hand, the direct methods avoid having
to make any estimate of the way that basin hurricanes relate to landfalling
hurricanes. On the other, there are more hurricanes in the basin than at landfall and so it
might be possible to predict basin numbers more accurately than landfalling numbers
(in some sense), and this accuracy might then feed through into
the landfall prediction.

In order to try and understand the relationship between these two methods a little better,
we investigate some of basic statistical
properties of the direct and indirect methods for predicting future hurricane rates.

In section~\ref{basics} we present some basic statistical ideas that we will use in our analysis.
In section~\ref{binomial} we set up the problem and derive expressions for the likely performance
of the indirect method in a general context.
In section~\ref{linear} we consider the performance of a set of simple prediction methods for
basin hurricane numbers.
In section~\ref{poisson} we specialize our analysis to the case where the basin hurricane numbers
are poisson distributed.
In section~\ref{mc} we perform some Monte-Carlo simulations
to check our approximations.
In sections~\ref{p1} and~\ref{p2} we apply the indirect method to make predictions
of the number of landfalling hurricanes, based on the basin hurricane number
predictions of~\citet{e01}.
Finally in section~\ref{conc} we discuss our results.

\section{Background on conditioning}\label{basics}

In this section we present some standard statistical results
that we will use later.

\subsection{Basic definitions~}
Consider two random variables $X$ and $Y$ with joint density
$f_{X,Y}$ and marginals $f_X$ and $f_Y$. The density of $Y|(X=x)$ is
defined as

\begin{equation}
f_{Y|X}(y|x) = \frac{f_{X,Y}(x,y)}{f_X(x)} \mbox{ where } f_X(x) \neq 0
\end{equation}

The conditional expectation is defined as $E(Y|X) = \psi(X)$ where

\begin{equation}
\psi(x) = E(Y|X=x) = \int_{R} y f_{Y|X}(y|x) dy.
\end{equation}

The conditional variance is defined as $\mbox{var}(Y|X) = \nu(X)$ where

\begin{equation}
\nu(x) = \mbox{var}(Y|X=x) = \int_{R} [y - E(Y|X=x)]^2 f_{Y|X}(y|x) dy.
\end{equation}

\subsection{Disaggregation of the variance}

From the definitions given above one can
derive a useful expression that disaggregates the variance
of $Y$ into conditional expectations and variances.

\begin{equation}\label{varevar}
\mbox{var}(Y|X) = E(Y^2|X) - [E(Y|X)]^2,
\end{equation}

and

\begin{equation} \label{condvar:eq}
\mbox{var}(Y) = E[\mbox{var}(Y|X)] + \mbox{var}[E(Y|X)].
\end{equation}

\subsection{Disaggregation of the variance of a product}

From equation~\ref{condvar:eq} we can then derive a useful method for
disaggregating the variance of a product.

First, it is always true that

\begin{equation}
\begin{array}{ll}
\mbox{var}(XY) & = E[\mbox{var}(XY|X)] + \mbox{var}[E(XY|X)]\\
& = E[X^2 \mbox{var}(Y|X)] + \mbox{var}[X E(Y|X)].
\end{array}
\end{equation}

Now, if $X$ and $Y$ are independent we have $E(Y|X) = E(Y)$ and
$\mbox{var}(Y|X) = \mbox{var}(Y)$ so

\begin{equation}
\begin{array}{ll}
\mbox{var}(XY) &=  E(X^2) \mbox{var}(Y) + E(Y)^2 \mbox{var}(X) \\
&= \mbox{var}(X) \mbox{var}(Y) +E(X)^2 \mbox{var}(Y) + E(Y)^2 \mbox{var}(X).
\end{array}
\end{equation}

We will use these expressions below.

\section{Basics of the conditional binomial model}\label{binomial}

We now set up our model. Overall our approach is to start with a very general mathematical
framework (e.g. we don't initially assume that hurricane numbers are
poisson distributed), derive what we can with this level of generality,
and make additional assumptions on the way through as and when necessary.

First, we need random variables for the
annual numbers of hurricanes in the basin and at landfall, and their
historical totals. We define these as follows:

\begin{itemize}

    \item Let $\{X_t:t=1,\dots,n\}$ be the sequence of annual historical hurricane numbers and
let $X=\sum_{t=1}^n X_t$.

    \item Let $\{Y_t:t=1,\dots,n\}$ be the sequence of annual historical \emph{landfalling} hurricane numbers
and let $Y=\sum_{t=1}^n Y_t$.

\end{itemize}

Now we consider estimating the proportion of hurricanes that make landfall, and the properties of the
most obvious estimator of that proportion. To start with, we don't assume that the number of hurricanes
in the basin is poisson, but we do assume that the probability of hurricanes making landfall is constant
in time, and is the same for all hurricanes. We write this (unknown) probability as $p$.
Then the number of hurricanes that make landfall in a given
year, given the number in the basin, is given by a binomial distribution:
\begin{equation}
Y_t|X_t \sim \mbox{binomial}(X_t,p)
\end{equation}

A useful analogy is that each basin hurricane is a coin toss, with a probability $p$ of giving a head.
The number of hurricanes
making landfall $Y_t$ is the number of heads in $X_t$ tosses.

Extending this to the total number making landfall over $n$ years, we also get a binomial:
\begin{equation}
Y|X_1,\dots,X_n \sim \mbox{binomial}(X,p)
\end{equation}

\subsection{Estimating the landfall proportion}

The most obvious way to try and estimate $p$ from the historical data is using the simple ratio
of the total number of historical landfalls to the total number of basin hurricanes:
\begin{equation}
\hat{p} = Y / X
\end{equation}

What are the properties of this estimator?
Is it unbiased, and what is the variance?

Wrt bias, first we note that:
\begin{equation}
E(\hat{p}|X_1,\dots,X_n) = p
\end{equation}

and that

\begin{eqnarray}
E(\hat{p})&=&E(E(\hat{p}|X_1,\dots,X_n))\\
          &=&E(p)\\
          &=&p
\end{eqnarray}

and we see $\hat{p}$ is unbiased.

Wrt variance, a standard result for the binomial distribution is that:

\begin{equation}
\mbox{var}(\hat{p}|X_1,\ldots,X_n) = p(1-p)/X
\end{equation}

Using equation~\ref{condvar:eq}, we can then decompose $\mbox{var}(\hat{p})$ as follows:
\begin{eqnarray} \label{varphat:eq}
\mbox{var}(\hat{p}) &=& E[\mbox{var}(\hat{p}|X_1,\ldots,X_n)] + \mbox{var}[E(\hat{p}|X_1,\ldots,X_n)]\\
                    &=& p(1-p) E(1/X).\label{varhat}
\end{eqnarray}

That is, the variance of the estimate of the proportion is given only in terms of the proportion itself
and $E(1/X)$. The proportion can be estimated using a plug-in estimator, but the $E(1/X)$ factor
is slightly harder to deal with, and can only be evaluated once we have settled on a distribution for $X$.
We consider this for the poisson distribution in section~\ref{poisson} below.

\subsection{Landfall predictions}

Now we consider making predictions of future landfalling hurricane numbers using the estimated proportion $\hat{p}$,
and a prediction of the mean number of basin hurricanes, which we write as $\mu = E(X_{n+1})$.
The first question is then how to estimate $\mu$.
One fairly general class of methods for estimating $\mu$ is to use
the historical data for the basin number of hurricanes in some way.
We can write this as
$\hat{\mu} = g(X_1,\ldots,X_n)$, where $g$ could be a linear or non-linear function
of the historical data.

The most obvious reasonable
forecast for the number of hurricanes making landfall is then
$\hat{p} \hat{\mu}$. What are the properties of this particular method?

We can establish the properties of this
predictor as follows.

For the bias:
\begin{eqnarray}
E(\hat{p} \hat{\mu}) &=& E(E(\hat{p} \hat{\mu}|X_1,\ldots,X_n))\\
                     &=& E(\hat{\mu}E(\hat{p}|X_1,\ldots,X_n))\\
                     &=& p E(\hat{\mu})\label{epmu:eq}
\end{eqnarray}

Note that if $\hat{\mu}$ is unbiased for $E(X_{n+1})$ then
equation~\ref{epmu:eq} implies that $\hat{p} \hat{\mu}$ is unbiased for
$E(Y_{n+1})$ (this is a stronger result than asymptotic
unbiasedness).

For the variance:
\begin{eqnarray}
\mbox{var}(\hat{p} \hat{\mu}) &=& E(\mbox{var}(\hat{p} \hat{\mu}|X_1,\ldots,X_n))
+ \mbox{var}(E(\hat{p} \hat{\mu}|X_1,\ldots,X_n)) \\
&=& E(\hat{\mu}^2 \mbox{var}(\hat{p} |X_1,\ldots,X_n))
+ \mbox{var}(\hat{\mu} E(\hat{p} |X_1,\ldots,X_n))  \\
&=& E(\hat{\mu}^2 p(1-p)/X) + p^2 \mbox{var}(\hat{\mu})  \\
&=& p(1-p) E(\hat{\mu}^2/X) + p^2 \mbox{var}(\hat{\mu}) \label{varpmu:eq}
\end{eqnarray}

We consider various approximations to this expression in the next two sections, which will
allow us to evaluate it in certain situations.

%
%

\section{Linear predictors of basin hurricane numbers}\label{linear}

We now move on to consider linear predictors of the number of hurricanes
in the basin i.e. methods that use a
weighted sum of historic values as an estimator of $\mu$.

We write this as:
\begin{equation}
\hat{\mu} = \sum_{i=1}^n w_i X_i.
\end{equation}

This linear framework includes the mixed baseline models of~\citet{j90},
and models that use linear regression of hurricane numbers on sea surface temperature.

To account for climate variability, the weights may be chosen to generate an estimator that uses only
recent data. For example:
\begin{equation}
w_i = \left\{\begin{array}{ll} 0, & \mbox{for} \ i=1,\ldots,n-m, \\
\frac1m, & \mbox{for } \ i=n-m+1,\ldots,n.
\end{array}\right.
\end{equation}

Under this model it \emph{may}, in some cases, be reasonable
to suppose that $\hat{\mu}$ is generated so that $\mbox{cov}(\hat{\mu}^2,
1/X)$ is small relative to $E(\hat{\mu}^2)E(1/X)$. Roughly speaking, this occurs if the errors we make
when estimating the proportion are not highly correlated with the errors we make
when making the basin prediction.

If we can assume that the covariance term is small then
 we can make some useful simplifications to equation~\ref{varpmu:eq}, as follows:
\begin{eqnarray}
\mbox{var}(\hat{p} \hat{\mu})
 &=& p(1-p) E(\hat{\mu}^2/X) + p^2 \mbox{var}(\hat{\mu}) \\
 &= &p(1-p)[E(\hat{\mu}^2)E(1/X) + \mbox{cov}(\hat{\mu}^2, 1/X)] + p^2 \mbox{var}(\hat{\mu}) \\
 & \approx & p(1-p)E(\hat{\mu}^2)E(1/X) + p^2 \mbox{var}(\hat{\mu})
\label{varpmuapprox:eq}
\end{eqnarray}

\section{Poisson model for basin hurricanes}\label{poisson}

We now specialize our analysis to the case where the number of hurricanes in
the basin can be modelled as a poisson distribution, which allows to approximate
the $E(1/X)$ term, and hence evaluate equations~\ref{varhat} and~\ref{varpmuapprox:eq}.

We start by assuming that the annual counts are poisson distributed,
with the same poisson mean in each year:
\begin{equation}
X_t \sim \mbox{poisson}(\mu) \mbox{ for all $t$}
\end{equation}

Then the total number of hurricanes over $n$ years is also poisson distributed:
\begin{equation}
X \sim \mbox{poisson}(n \mu)
\end{equation}

(statisticians usually prove this by inspection of moment generating functions).

At this point we briefly mention a small mathematical problem, which is that
we are now going to consider $1/X$, even though $X$, being poisson distributed,
can take values of $0$. To get around this problem rigourously one can
condition on $X>0$, which would introduce a small
adjustment factor to the expressions derived below.
We will, however, ignore this. Effectively we are assuming that the probability of $X$
being zero is small, and this should be borne in mind when applying the results we derive.
This should be a reasonable assumption if $X$ is the number of Atlantic basin hurricanes, but
would not reasonable if $X$ we the number of category 5 Atlantic basin hurricanes, for instance.

Our approximation for $E(1/X)$ is based on a Taylor expansion for the annual numbers:
\begin{eqnarray}
& E(1/X_t) = \frac{1}{\mu}\left[1+ \frac{1}{\mu} + 2\frac{1}{\mu^2}
+ O(\frac{1}{\mu^3})\right] \\
\Rightarrow & E(1/X)  = \frac{1}{n \mu}\left[1+ \frac{1}{n \mu} +
2\frac{1}{n^2 \mu^2} + O(\frac{1}{n^3 \mu^3})\right]
\label{e1overX:eq}
\end{eqnarray}

Thus, to first order, $E(1/X) \approx \frac{1}{n \mu}$.

If we take this first order approximation and substitute it
into equation~\ref{varhat} then we get:
\begin{equation}
\mbox{var}(\hat{p}) \approx \frac{p(1-p)}{n \mu},
\end{equation}

And if we substitute it into equation~\ref{varpmuapprox:eq} we get
\begin{equation} \label{varpmuapprox2:eq}
\mbox{var}(\hat{p} \hat{\mu}) \approx \frac{p(1-p) E(\hat{\mu}^2)}{n \mu}
+ p^2 \mbox{var}(\hat{\mu}).
\end{equation}

One simple prediction method for the mean number of hurricanes in the basin
is to take a straight average of $m$ years of data.
Given this,
\begin{equation}
\mbox{var}(\hat{\mu}) = \mu /m \label{varmuhat}
\end{equation}
and
\begin{eqnarray}
E(\hat{\mu}^2) &=& \mu/m + \mu^2\\
               &=& \mu (1+ m \mu) / m
\end{eqnarray}

In this case we get:
\begin{equation}\label{answer}
\mbox{var}(\hat{p} \hat{\mu}) \approx \frac{p(1-p)(1+m\mu)}{nm} + p^2
\frac{\mu}{m}.
\end{equation}

How accurate are these results based on the first-order approximations?
They will be reasonable if $n$ is large. Better
approximations to $\mbox{var}(\hat{p})$ and $\mbox{var}(\hat{p} \hat{\mu})$
can easily be generated by using higher order terms in the approximation of
$E(1/X)$.

\section{Simulation tests}\label{mc}

We now test the first order approximation using Monte-Carlo simulations.
We consider the following situation:
\begin{itemize}

    \item We estimate the mean number of hurricanes making landfalling using just the
    last 11 years of landfalling data. This is one of our predictions.

    \item We estimate the mean number of \emph{basin} hurricanes using the same 11 years of data

    \item We convert the basin estimate to an estimate for landfalling numbers using
    an estimated proportion, which is based on between 11 and over 50 years of data.
    11 of the years of data used to estimate the proportion are the same data that is
    used to estimate the rates.

    \item We estimate the variances of all these predictions

\end{itemize}

\begin{figure}[!hb]
  \begin{center}
    \scalebox{0.7}{\includegraphics{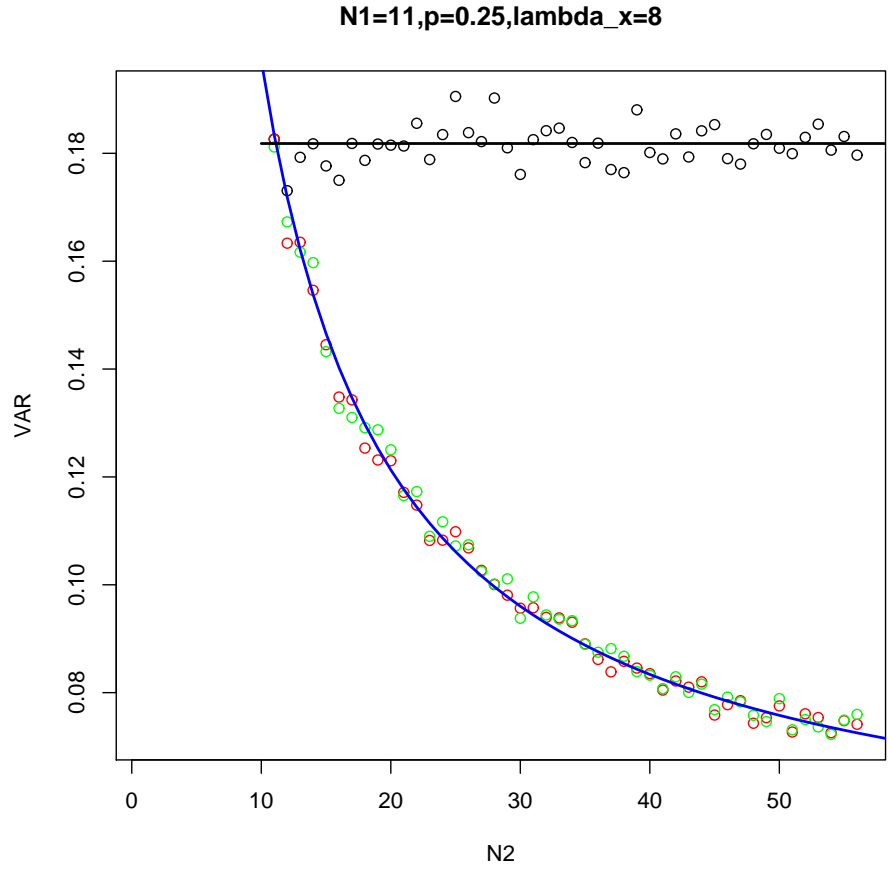}}
  \end{center}
    \caption{
Variances from analytic expressions and Monte Carlo simulations.
The black line shows the variance of the direct landfall prediction, based
on 11 years of data. The grey line shows an estimate of the variance of the indirect landfall prediction,
based on 11 years of basin data, and $N2$ years of basin data, using equation~\ref{answer}.
The black circles show simulation-based estimates of the variance of the direct prediction, and the grey
circles show simulation-based estimates of the variance of the indirect prediction.
The simulations validate the approximations used to derive equation~\ref{answer}.
}
     \label{f01}
\end{figure}

Using Monte-Carlo simulations we can compare the variance estimate given by equation~\ref{answer}
with the real variance estimates. The results are given in figure~\ref{f01}.
The black-line gives the variance of the landfall prediction based on 11 years of historical
landfall data, from equation~\ref{varmuhat}. The black-dots give estimates of this variance
based on the simulations. The blue-line gives our theoretical approximation to the
variance from the indirect method, based on equation~\ref{answer}. The coloured dots give estimates
of the variance from the indirect method based on the simulations. We see that:

\begin{itemize}

    \item The theoretical estimate of the variance for the indirect method is in very good
    agreement with the results from the simulations, even though we've only used a
    first order approximation to derive equation~\ref{answer}.

    \item The variance of the indirect method is lower than the variance of the direct
    method when the proportion is estimated using more years of data than are being
    used for the rate estimates. Using 35 or more years of data makes the indirect
    method more than twice as accurate, in terms of variance.

\end{itemize}

\section{Applying the indirect method}\label{p1}

We now make some predictions of future numbers of landfalling hurricane numbers
by converting the basin hurricane number predictions given in~\citet{e01} to
landfalling predictions.

\subsection{Step 1: predicting numbers of basin hurricanes}

The predictions of numbers of basin hurricanes that we use are taken
from~\citet{e01},
in which mixed baseline models are used to predict future
numbers of hurricanes in the basin. These models are based on an
analysis of change-points in the historical time-series of
hurricane numbers. The intervals between change-points are taken
as periods of levels of constant hurricane activity, and future
activity is then predicted on the assumption that the current
level of activity will continue. The prediction is given by an
optimal combination of the observed activity rates in the
historical data, where `optimal' is defined as minimising
mean-square-error, and trades off the need to use as much of the
historical data as possible (for increased accuracy) against the
desire to use only recent data (because it is likely to be the
most relevant for the future).

The predictions from~\citet{e01} are based on the change-point
analysis of~\citet{elsnerj00} and~\citet{e02a}. We include predictions based on both of these
change-point analyses to get an idea of the level of sensitivity
of the results to the details of the methods used to detect the
change-points.

\subsection{Step 2: relating basin hurricane numbers to landfalling hurricane numbers}
The empirical relationships we use to convert the number of basin hurricanes
to a number of landfalling hurricanes are simple estimates of the probability that hurricanes will
make landfall, based on historical hurricane data for 1950 to
2005. For cat 1-5 hurricanes, we estimate this probability to be
0.254 (with a standard error of 0.058). For cat 3-5 hurricanes, we
estimate this probability to be 0.240 (with a standard error of
0.057).

\subsection{Step 3: converting basin predictions to landfalling predictions}

The predictions used in step 1 above produce estimates for the mean
number of basin hurricanes, the variance of the number of basin
hurricanes, and the standard error on the mean. The empirical
relationships in step 2 tell us how to convert the number of
hurricanes in the basin into the number at landfall, given
information about the number in the basin. How, then, should we
combine this information to tell us about the distribution of the
number of hurricanes at landfall? A complete solution for the
distribution of the number of hurricanes at landfall would be
slightly complicated to derive. The mixed baseline models
themselves don't give a probabilistic prediction, but just the
first two moments and the standard error on the mean. Although
they are built on the assumption that the number of hurricanes is
poisson distributed, the predictions they produce cannot strictly
be interpreted as poisson distributions because the mean and
variance are not equal. However, deriving expressions for the
mean, variance, and standard error on the mean for the number of
landfalling hurricanes, which is all we are interested in, is
rather simple. These expressions are given in~\citet{e06}.

Putting all of this together, we make predictions for:
\begin{itemize}
    \item The number of landfalling cat 1-5 hurricanes, based on the basin number of cat 1-5 hurricanes
    \item The number of landfalling cat 3-5 hurricanes, based on the basin number of cat 3-5 hurricanes
\end{itemize}

In each case we predict the mean number of hurricanes, the
variance of the number of hurricanes, and the standard error on
the mean (which is based on both the standard error on the
prediction of the basin number of hurricanes, and the standard
error of the estimate of the proportion making landfall).

\section{Predictions from the indirect method}\label{p2}

\subsection{Predictions based on Elsner change points}

The results from our analysis based on the change-points
from~\citet{elsnerj00} are shown in table 1. The first four rows
of this table are for cat 1-5 storms, while the second four rows
are for cat 3-5 storms. As an example, consider the first row.
From~\citet{e01}, table 5, we can see that the short baseline
model predicts 8.45 hurricanes in the basin, with a standard error
of 0.877. Converting that to a prediction of landfalling
hurricanes using the estimated probability of landfall of 0.254
gives 2.15 hurricanes, which is the value for the mean shown in
the first row of table 1. Similarly the variance of the number of
landfalls in this case is 2.74. The standard error, which arises
because of (a) the uncertainty in the prediction of the number of
storms in the basin and (b) the uncertainty in how to convert that
number to a prediction at landfall, is 0.549.

How do these new predictions for the mean number of landfalling
hurricanes compare with the previous results in~\citet{e01}?
Considering the most complex model in each case (model 4 in table
1), the prediction for cat 1-5 storms changes from 2.09 to 2.08,
which is insignificant. For cat 3-5 (model 8 in table 1), however,
the prediction changes from 0.82 to 0.92. This is a more
significant change (although still well within the standard error
estimates). What is driving this increase? It turns out that the
percentage increase in the basin number of severe storms that we
have seen in the last 11 years is rather larger than the
percentage increase in the number of severe storms at landfall
(basin severe storms numbers have increased by 88\% relative to
the long-term baseline, while landfalling severe storms have only
increased by 40\%). This increase in the basin severe storms leads
to a high prediction of the future number of severe storms in the
basin, and this in turn leads to a high prediction of the number
of severe storms at landfall when using this method of predicting
landfall numbers from the predicted basin numbers. As discussed in
the introduction, there are good reasons to think this might be a
more accurate prediction than the lower prediction based on the
landfall data alone, since the landfall data is so sparse.

\subsection{Predictions based on RMS change points}

The results from our analysis based on the change-points
from~\citet{e03} are shown in table 2. Once again, the first four
rows of this table are for cat 1-5 storms, while the second four
rows are for cat 3-5 storms.

We see a small decrease in the prediction of the number of cat 1-5
storms, and another, although smaller, increase in the number of
cat 3-5 storms.

\section{Conclusions}\label{conc}

One possible way to predict landfalling hurricane numbers is to first predict
basin hurricane numbers and then convert the basin numbers to landfall using
an estimate of the proportion of the basin hurricanes that make landfall. This method
can be compared with the simpler method of just predicting landfall numbers directly.
We have performed some statistical analysis of these methods, to try and understand
which is likely to be more accurate.
In particular we have considered a situation where the direct method consists
of estimating the landfall rates using an 11 year average of historical landfalling rates,
and the indirect method consists
of estimating basin rates using an 11 year average and then converting that to landfall
rates using a proportion based on more than 11 years of data. Assuming that the probability of
individual hurricanes making landfall is constant in time then we have shown that the indirect method
is more accurate, and the more data is used to estimate the proportion, the more accurate
it becomes relative to the indirect method. Furthermore we have derived expressions for
the variance of the indirect method, and using simulations have shown that a simple
analytic expression for the variance of the indirect method works well.

We then apply the indirect method to convert some previous predictions of basin hurricane
numbers into predictions of numbers of landfalls.
The results for landfalling cat 1-5 storms are not that different
between this method and results from predicting landfalling storm
numbers directly from historical landfalls. The results for cat
3-5 storms, however, show higher predictions. This is because the
number of cat 3-5 storms in the basin has increased more in recent
years (proportionately) than the number of cat 3-5 storms at
landfall.

Preliminary results (as yet unpublished) suggest that the
hypothesis that the probability of storms making landfall doesn't
change in time cannot be rejected. This lends weight to the idea
that these higher predictions of future numbers of intense
landfalling storms may be more reliable. However, the difference
between the two predictions is well within the standard error
estimates, and so either prediction could easily have been much
higher or lower just due to random effects.

\bibliography{arxiv}

\begin{thebibliography}{6}
\providecommand{\natexlab}[1]{#1}
\providecommand{\url}[1]{\texttt{#1}}
\expandafter\ifx\csname urlstyle\endcsname\relax
  \providecommand{\doi}[1]{doi: #1}\else
  \providecommand{\doi}{doi: \begingroup \urlstyle{rm}\Url}\fi

\bibitem[Binter et~al.(2006)Binter, Jewson, Khare, O'Shay, and Penzer]{e01}
R~Binter, S~Jewson, S~Khare, A~O'Shay, and J~Penzer.
\newblock {Year ahead prediction of US landfalling hurricane numbers: the
  optimal combination of multiple levels of activity since 1900}.
\newblock \emph{arXiv:physics/0611070}, 2006.
\newblock RMS Internal Report E01.

\bibitem[Elsner et~al.(2000)Elsner, Jagger, and Niu]{elsnerj00}
J~Elsner, T~Jagger, and X~Niu.
\newblock {Changes in the rates of North Atlantic major hurricane activity
  during the 20th Century}.
\newblock \emph{Geophysical Research Letters}, 27:\penalty0 1743--1746, 2000.

\bibitem[Jewson(2007)]{e06}
S~Jewson.
\newblock {Predicting Hurricane Numbers from Sea Surface Temperature: closed
  form expressions for the mean, variance and standard error of the number of
  hurricanes}.
\newblock \emph{arXiv:physics/0701167}, 2007.
\newblock RMS Internal Report E06.

\bibitem[Jewson et~al.(2005)Jewson, Casey, and Penzer]{j90}
S~Jewson, C~Casey, and J~Penzer.
\newblock {Year ahead prediction of US landfalling hurricane numbers: the
  optimal combination of long and short baselines}.
\newblock \emph{arxiv:physics/0512113}, 2005.

\bibitem[Jewson and Penzer(2006)]{e02a}
S~Jewson and J~Penzer.
\newblock {An objective change point analysis of historical Atlantic hurricane
  numbers}.
\newblock \emph{arXiv:physics/0611071}, 2006.
\newblock RMS Internal Report E02a.

\bibitem[O'Shay and Jewson(2007)]{e03}
A~O'Shay and S~Jewson.
\newblock {Year ahead prediction of US landfalling hurricane numbers: the
  optimal combination of multiple levels of activity since 1900: sensitivity to
  alternative change point definitions}.
\newblock \emph{arXiv:physics/0701227}, 2007.
\newblock RMS Internal Report E03.

\end{thebibliography}

\newpage

\newpage
 \begin{table}[h!]
   \centering
 \begin{tabular}{|c|c|c|c|c|c|c|c|}
  \hline
  1 & 2 & 3 & 4 & 5 & 6 & 7 & 8\\
  \hline
 Model & Model Name(No. Yrs)& Basin Model &            LF Model & Mean & Var & Var/Mean & RMSE \\
  \hline
                                      1& SBL (11)& Basin Cat15 & LF Cat15 &  2.15&  2.74&  1.28& 0.549  \\
                            2& 2 active pds (33) & Basin Cat15 & LF Cat15 &  1.75&  2.47&  1.41& 0.492  \\
         3& 2 pds, both active opt wi (106) & Basin Cat15      & LF Cat15 &  2.08&  2.18&  1.05& 0.532  \\
                                  4& 4 pds (106) & Basin Cat15 & LF Cat15 &  2.08&  2.18&  1.05& 0.532  \\
  \hline
                                      5& SBL (11)& Basin Cat35 & LF Cat35 &  0.98&  1.24&  1.26& 0.315  \\
                            6& 2 active pds (33) & Basin Cat35 & LF Cat35 &  0.87&  1.13&  1.29& 0.294  \\
          7& 2 pds, both active opt wi (106) & Basin Cat35     & LF Cat35 &  0.92&  0.99&  1.08& 0.296  \\
                                  8& 4 pds (106) & Basin Cat35 & LF Cat35 &  0.92&  0.99&  1.08& 0.296  \\
 \hline
 \end{tabular}
 \caption{
 Predictions for landfalling hurricane numbers, based on the
 change points from~\citet{elsnerj00}, and the method described in
 the text.
 }
 \end{table}

 \begin{table}[h!]
   \centering
 \begin{tabular}{|c|c|c|c|c|c|c|c|}
  \hline
  1 & 2 & 3 & 4 & 5 & 6 & 7 & 8\\
  \hline
 Model & Model Name(No. Yrs)& Basin Model &            LF Model & Mean & Var & Var/Mean & RMSE \\
  \hline
                                      1& SBL (11)& Basin Cat15 & LF Cat15 &  2.15&  2.20&  1.02& 0.549  \\
                            2& 2 active pds (33) & Basin Cat15 & LF Cat15 &  1.82&  1.94&  1.06& 0.499  \\
         3& 2 pds, both active opt wi (106) & Basin Cat15      & LF Cat15 &  2.05&  2.09&  1.02& 0.526  \\
                                  4& 5 pds (106) & Basin Cat15 & LF Cat15 &  2.05&  2.09&  1.02& 0.526  \\
  \hline
                                      5& SBL (11)& Basin Cat35 & LF Cat35 &  0.98&  1.00&  1.02& 0.315  \\
                            6& 2 active pds (33) & Basin Cat35 & LF Cat35 &  0.87&  0.89&  1.02& 0.294  \\
          7& 2 pds, both active opt wi (106) & Basin Cat35     & LF Cat35 &  0.91&  0.93&  1.02& 0.295  \\
                                  8& 5 pds (106) & Basin Cat35 & LF Cat35 &  0.91&  0.93&  1.02& 0.295  \\
 \hline
 \end{tabular}
 \caption{
 Predictions for landfalling hurricane numbers, based on the
 change points from~\citet{e01}, and the method described in
 the text.
 }
 \end{table}

\end{document}